\documentclass{osa-article}

\journal{oe}

\begin{document}

\title{Edge detection based on joint iteration ghost imaging}

\author{Cheng Zhou,\authormark{1,2,3} Gangcheng Wang,\authormark{1}  Heyan Huang,\authormark{4} Lijun Song,\authormark{2,3,5} Kang Xue,\authormark{1,6} }

\address{\authormark{1}Center for Quantum Sciences and School of Physics, Northeast Normal University, Changchun 130024, China\\
\authormark{2}Institute for Interdisciplinary Quantum Information Technology and School of Information Engineering, Jilin Engineering Normal University, Changchun 130052, China\\
\authormark{3}Jilin Engineering Laboratory for Quantum Information Technology, Changchun 130052, China\\
\authormark{4}College of Science, Shanghai Institute of Technology, shanghai 201418,  China\\
\authormark{5}ccdxslj@126.com\\
\authormark{6}xuekang@nenu.edu.cn\\}



\begin{abstract}
Imaging and edge detection have been widely applied and played an important role in security checking and medical diagnosis. However, as we know, most edge detection based on ghost imaging system require a large measurement times and the target object image cannot be provided directly. In this work, a new edge detection based on joint iteration of projected Landweber iteration regularization and guided filter ghost imaging method have been proposed which can be improved the feature detection quality in ghost imaging. This method can also achieve high quality imaging. Simulation and experiment results show that the spatial information and edge information of target object are successfully recovered from the random speckle patterns without special coding under a low measurement times, and the edge image quality is improved remarkably. This approach improves the the applicability of ghost imaging, and can satisfy the practical application fields of imaging and edge detection at the same time.
\end{abstract}

\section{Introduction}

Ghost imaging (GI) is a novel optical imaging technology that is rather different from conventional modalities. In conventional optical imaging, the object image is directly acquired by a multi-pixel detector. Surprisingly, GI uses a single-pixel detector to detect the total light signal strength of object, while a detector with spatial resolution measures the information about the light source. The image can be retrieved merely by correlating the signals of these two detectors, but not either one alone \cite{p1995}.  Remarkably, GI has the superiority of higher sensitivity in detection and anti-atmospheric disturbance than conventional optical imaging \cite{g2004}. Hence, GI is increasingly focused on its applications, such as remote sensing \cite{rs2012,rs2015,rs2016}, optical coherence tomography \cite{oct2012,oct2018}.

In 2008, a computational ghost imaging (CGI) theoretical scheme that requires only one single-pixel detector was proposed by Shapiro \cite{tcgi}. Subsequently, CGI scheme was experimental demonstrated by Bromberg \cite{ecgi}. Then, many methods was proposed to improve the imaging quality, including compressive GI \cite{csgi}, differential GI \cite{dgi}, pseudo-inverse GI \cite{tpgi,epgi} and so on \cite{hgi,igi,ngi,npgi}. Especially, compressive GI can achieve the high-quality reconstruction image under undersampling \cite{csgi}, which promotes the practical application of GI technology  \cite{tdgi,gil,gt,gc,ftxgi,exgi,lsxgi,txgi,elgi}.

Recently, edge information detection of target object for GI was considered \cite{ggi,ssgi,spsgi,sigi,spigi,eedgi,cgei}. Edge detection is widely used in computer vision, target recognition, earth observation and security check \cite{ed1,ed2}. As we all know, traditional edge detection methods (e.g. Canny \cite{canny}, Sobel \cite{sobel}, Roberts \cite{robert}, etc.) rely on the original image. However, in many practical application scenarios in which harsh or noisy environments, the traditional edge detection methods are ineffective because the image information of the target object is difficult to obtain. Different from traditional edge detection methods, edge detection based on GI scheme can detect edge information directly without needing the original image. Hence, edge detection based on GI can solve the problem of disturbance due to its advantages on good anti-disturbance imaging and direct edge detection of unknown objects.

 Here, some edge detection based on GI methods have been reviewed \cite{ggi,ssgi,spsgi,sigi,spigi,eedgi,cgei}. Liu et al. proposed a gradient GI (GGI) scheme which directly achieved the edge information of an unknown target object \cite{ggi}. Subsequently, a more optimized edge detection method named speckle-shifting GI (SSGI) was reported by Mao \cite{ssgi}. The SSGI scheme does't need the gradient angle or any other prior knowledge of the object in GGI. Then, wang et al. \cite{spsgi} proposed another similar method called subpixel-speckle-shifting GI (SPSGI), which is based on a set of subpixel-shifted Walsh-Hadamard speckle pattern pairs and has the advantage of enhancing the resolutions of the edge detection. Meanwhile, Yuan et al. \cite{sigi} used structured illuminations based on the interference principle to get edge information, and the method can extract the edges of binary and gray targets in any direction at the same time. From the perspective of light field coding, special sinusoidal patterns for the x-direction edge and also y-direction edge of the unknown object were designed by Ren \cite{spigi}. Furthermore, a novel variable size Sobel operator whose coefficients are isotropic and sensitive to all directions was designed and used for edge detection based on GI by Ren et al, whereby the edge detection based on GI method could directly achieve the edges of an unknown object without choosing the gradient angle or any other prior knowledge of the object \cite{eedgi}. However, these methods still have some shortcomings, such as high measurement times and poor quality of edge information acquisition.

In order to improve the efficiency of edge detection based on ghost imaging, Guo et al.  proposed a compressed ghost edge imaging (CGEI) scheme, which designed special random patterns with the characteristic of different speckle-shifting, and used compressed sensing technology and Sobel operator, whereby the measurements required for edge detection can be further reduced \cite{cgei}. Noteworthy, these methods can only obtain edge information singly, unless the reconstruction calculation of the whole information is carried out again. If the edge information and the whole image information of the object can be obtained simultaneously at a lower sampling times, this will greatly promote the air surveillance and ocean monitoring application of GI. We find that the problem can be well solved by using the compressive GI based on guided filtering method \cite{lggi}.

In this paper, we demonstrate an edge detection based on joint iteration ghost imaging (JIGI) method for simultaneously acquiring the global edge and whole image information. Because the JIGI method is based on projected Landweber iteration regularization and guided filter, in which guided filter is an edge-preserving filter which can enhance the signal-to-noise ratio of edge detection, the proposed method have some benefits: 1) High efficiency: simultaneous acquisition of high quality edge and whole image information is implemented with fewer measurement times; 2) More convenient: edge detection can be realized in any light field without designing special light field and pair measurement; 3) Strong universality: it is not limited to computational ghost imaging using light field modulation equipment, but also suitable for other ghost imaging methods such as dual-path pseudo-thermal ghost imaging.

\section{Theoretical analysis}
In CGI system, the detection light source is generated from a light beam through a spatial light modulator (SLM) [or a digital micromirror devices (DMD)], and then passes through an optical lens to adjust the size of light beam. The transmitted or reflected light field $I^{(m)}(i,j)$ ($m=1,2,3,\cdots,M$ represents the number of measurement times) passing through the target object with a transmission coefficient of $O(i,j)$ is recorded by a single-pixel detector; and the detection value obtained from the $m$-th sampling is expressed as $B^{(m)}$.


Here, by reconfiguring the elements of each speckle pattern (dimensions $r\times c$) pre-generated by computer into a row vector of length $K=r\times c$ to form one row of the matrix $\Phi$, we obtain the following $M\times K$ matrix, based on $A$ measurements:

\begin{equation}
A=\left[
             \begin{array}{cccc}
               I_1(1,1) & I_1(1,2) & \cdots & I_1(r,c) \\
               I_2(1,1) & I_2(1,2) & \cdots & I_2(r,c) \\
               \vdots & \vdots & \ddots & \vdots \\
               I_M(1,1) & I_M(1,2) & \cdots & I_M(r,c) \\
             \end{array}
           \right],
\end{equation}

The $M$ results from the single-pixel detector can be permutated into an $M\times 1$ column vector $y$:
\begin{equation}
y=\left[B^{(1)},B^{(2)},\cdots,B^{(M)}\right]^T,
\end{equation}

 Then, if we denote the unknown target object $O(i,j)$ as an $K$ dimensional column vector $x$ $(K\times 1)$, we will have the framework $y=Ax$, and the matrix form is expressed as:

 \begin{equation}\label{cs1}
   \left[
   \begin{array}{c}
               B^{(1)} \\
               B^{(2)} \\
               \vdots  \\
               B^{(M)} \\
             \end{array}
   \right]= \left[
             \begin{array}{cccc}
               I^{(1)}(1,1) & I^{(1)}(1,2) & \cdots & I^{(1)}(r,c) \\
               I^{(2)}(1,1) & I^{(2)}(1,2) & \cdots & I^{(2)}(r,c) \\
               \vdots & \vdots & \ddots & \vdots \\
               I^{(M)}(1,1) & I^{(M)}(1,2) & \cdots & I^{(M)}(r,c) \\
             \end{array}
             \right] \left[
                        \begin{array}{c}
                           O(1,1) \\
                           O(1,2) \\
                           \vdots \\
                           O(r,c) \\
                        \end{array}
                      \right].
\end{equation}

\subsection{Step 1: projected Landweber iteration regularization}
As we all know, some special regularization methods are exploited to solve Eq.(\ref{cs1}), such as projected Landweber iteration regularization (PLIR) \cite{lggi,landweber1,landweber2}. Here, we use PLIR to get the preliminary reconstructed image. The PLIR is defined as\cite{landweber2}:

\begin{equation}\label{land}
x_{t} = x_{t-1}+\omega PA^T(y-Ax_{t-1}), ~~~~~~t=1,2,3,\cdots,
\end{equation}
where $P$ is the pseudo-inverse of $A^T A$, $\omega$ is the gain factor to control the convergence speed, $x_{t}$ is the approximate solution of Eq.(\ref{cs1}), and $x_{t-1}$ is the approximate solution of the previous (initial supposition: $x_0=\left[0,0,\cdots,0\right]^T$). Let $x_{t-1} := x_0$, and then we first obtain initial reconstruction image $x_1$.

\subsection{Step 2: guided filter}
In order to realize the edge detection based on GI, the resulting image $x_1$ is reshaped $K\times 1$ dimensions into a matrix of $r\times c$ dimensions and processed with a guided filter method. Here,  we denote the guided filter as \cite{guidedfilter1,guidedfilter2}:

\begin{equation}\label{gfold}
 q_t=\textrm{guidefilter}(I_t,x_t), ~~~~~~t=1,2,3,\cdots,
\end{equation}
where $x_t$ is the filter input image (i.e., the reconstruction result of PLIR), $I_t$ is the guidance image ($t=1:I_1=x_1; t>1:I_t=q_{t-1}$, $ q_t=I_t$ are the matrix of $r\times c$ dimensions ), $q_t$ is an output image. The filtering output at a pixel $i$ is expressed as

\begin{equation}
q_{ti}=\sum_j W_{i,j}(I_t)x_{tj},
\end{equation}
where $i$ and $j$ are pixel indexes. The filter kernel $W_{i,j}$ is a function of the guidance image $I$ and is independent of $x$, this definition follows\cite{guidedfilter1}
\begin{equation}
W_{i,j}(I)=\frac{1}{|\omega|^2} \sum_{k:(i,j)\in \omega_k} \left[1+\frac{(x'_i-\mu_k)(x'_j-\mu_k)}{\sigma_k^2+\epsilon}\right],
\end{equation}
where, $x'$ is the coordinate of the pixel, $\omega_k$ is the $k$-th kernel function window, $|\omega|$ is the number of pixels in $\omega_k$, $\epsilon$ is a regularization parameter. Here, $\mu_k$ and $\sigma_k^2$ are the mean and variance of $x$ in $\omega_k$.

In guided filter, it assumes that there is a local linear relationship between the guided image $I_t$ and the output image $q_{ti}$ in a window $\omega_k$ centered at the pixel $k$:

\begin{equation}\label{qab}
q_{ti}= a_kI_{ti}+b_k, \forall_i\in \omega_k,
\end{equation}
where $(a_k, b_k)$ are some linear coefficients assumed to be
constant in $\omega_k$. Let's take the gradient of both sides of Eq.~(\ref{qab}):

\begin{equation}\label{dqa}
\nabla q= a\nabla I.
\end{equation}

Such local linear model ensures that $q$ has an edge only if $I$ has an edge. This model [Eq.(\ref{dqa})] has been proven useful in image matting \cite{imagematting}, image super-resolution \cite{super-resolution}, and haze removal \cite{hazeremoval}. To determine the linear coefficients $(a_k, b_k)$, we minimize the following cost function in the window $\omega_k$:

\begin{equation} \label{edqa}
E(a_k, b_k)=\sum_{i\in \omega k}((a_kI_{ti}+b_k-x_{ti})^2+\epsilon a_k^2),
\end{equation}
where, $\epsilon$ is a regularization parameter penalizing large $a_k$. Using linear ridge regression method, the coefficient of Eq. (\ref{edqa}) are obtained as follows:

\begin{equation} \label{edqak}
a_k=\frac{\frac{1}{|\omega|}\sum_{i\in \omega_k}I_{ti}x_{ti}-\mu_k \bar{x}_k}{\sigma^2_k+\epsilon},
\end{equation}

\begin{equation} \label{edqbk}
b_k=\bar{x_t}_k-a_k\mu_k.
\end{equation}

Here, $\bar{x}_k=\frac{1}{|\omega|}\sum_{i\in \omega_k}x_{ti}$ is the mean of $x$ in $\omega_k$. Ordinary, we average all the possible $q_{ti}$ values as the final $q_{ti}$. Hence, after computing $(a_k,b_k)$ for all windows $\omega_k$ in the image, we compute the filtering output by

\begin{eqnarray}
q_{ti}&=&\frac{1}{|\omega|}\sum_{k:i\in \omega_k}(a_kI_{ti}+b_k),\\
      &=&\bar{a}_iI_{ti}+\bar{b}_i \label{qgfo},
\end{eqnarray}

where, $\bar{a}_i=\frac{1}{|\omega|}\sum_{k:i\in \omega_k}a_k,\bar{b}_i=\frac{1}{|\omega|}\sum_{k:i\in \omega_k}b_k$. In order to obtain the edge information of GI, we add $a_k$ to the output of Eq.(\ref{gfold}). Hence, the new guided filter that contains both global edge and whole image information of the object is expressed as:

\begin{equation}\label{ngfold}
 [q_t,a_k]=\textrm{guidefilter}(I_t,x_t), ~~~~~~t=1,2,3,\cdots.
\end{equation}

\subsection{Step 3: joint iteration}

\begin{figure}[htbp]
\centering
\includegraphics[width=12cm]{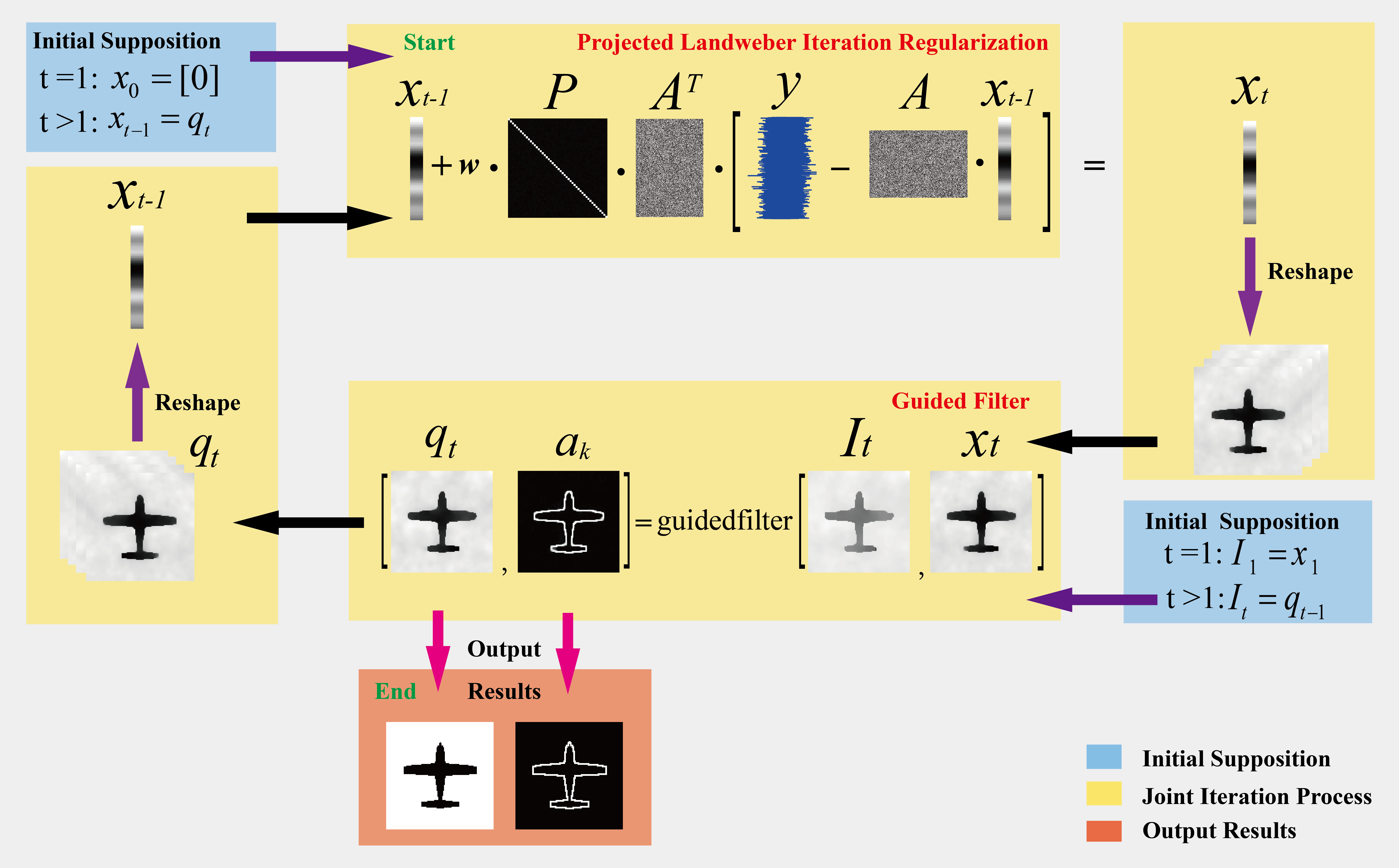}
\caption{(Color online) Schematic diagram of edge detection based on joint iteration ghost imaging.}
\label{methods}
\end{figure}

The output result $q_t$ of Eq.~(\ref{ngfold}) in Step 2 is taken as the input image $x_{t-1}$ of Eq.~(\ref{land}) in Step 1, i.e., $x_{t-1}=q_t$. Then, follow the loop iterations from step 1 to step 2 until the output results $q_t$ and $a_k$ converges at high quality. In this way, the edge detection based on JIGI method is realized by joint iteration of PLIR and guided filter, which obtains the high quality edge and whole image information at the same time. The joint iteration process is shown in Fig.\ref{methods}.

More specifically, in our edge detection based on joint iteration ghost imaging method, PLIR is used to constantly add limited information into the reconstruction results, which makes a small number of signals be fully utilized. Meanwhile, according to the result information of the PLIR and the detailed information of the guidance image (the previous guided filter output image), the whole image information and the feature information (edge information) of the image can be acquired. The mechanism for obtaining edge information through $a_k$ is explained in detail below. It is shown from the Eq.~(\ref{edqak}-\ref{edqbk}) that guided filter has the edge-preserving smoothing property \cite{guidedfilter2}. First, let $I_t=x_t$. Then we rewrite the Eq.~(\ref{edqak}) as:

\begin{equation} \label{edqakr}
a_k=\frac{\frac{1}{|\omega|}\sum_{i\in \omega_k}I_{ti}^2-\mu_k ^2 }{\sigma^2_k+\epsilon},
\end{equation}
where,

\begin{equation} \label{edqakrs}
\sigma^2_k=\frac{1}{|\omega|}\sum_{i\in \omega_k}I_{ti}^2-\mu_k ^2.
\end{equation}

Here, $\sigma^2_k$ denotes the variance of region $ \omega_k$, i.e., the local variance of the guidance image $I_{ti}$. So, Eq.~(\ref{edqakr}) can be further expressed as:

\begin{equation} \label{edqakrss}
a_k=\frac{\sigma^2_k}{\sigma^2_k+\epsilon}.
\end{equation}

This can be explained intuitively as follows. If the region $ \omega_k$ contains more texture and edge features, then $a_k$ becomes close to 1 on account of $\sigma^2_k$ has a larger value. Contrary, if the region $ \omega_k$ is constant or relatively smooth, then $a_k$ becomes close to 0 because the value of  $\sigma^2_k$ is very small. From the above Eq.~(\ref{edqakr}-\ref{edqakrss}) can see, $a_k$ is the global edge information of output image $q_{ti}$, and $b_k$ is the internal information (i.e., contains no edge information).

\subsection{Performance evaluation}
In order to objectively evaluate the performance of our edge detection method, the signal-to-ratio (SNR) is used, which is defined:

\begin{equation}\label{snr}
\textrm{SNR} = \frac{mean(q_{edge})-mean(q_{back})}{(var(q_{back}))^{0.5}},
\end{equation}
where $q_{edge}$ and $q_{back}$ are the intensities of the edge detection result in the object edge and background region respectively, and var stands for the variance. At the same time, we use a peak signal-to-ratio (PSNR) to estimate the quality of edge detection image. The PSNR reads:

\begin{equation}\label{psnr}
\textrm{PSNR} = 10\times log_{10}\left[ \frac{\textrm{max}Val^2}{MSE}\right],
\end{equation}
where $\textrm{max}Val^2$ is the maximum possible pixel value of the image and
\begin{equation}\label{MSE}
\rm{MSE}=\frac{1}{r\times c}\sum^{r}_{i=1}\sum^{c}_{j=1}\left[O_{edge}-a_k\right]^2,
\end{equation}
where $O_{edge}$ represents the original edge image consisting of $r\times c$ pixels, and $a_k$ denotes the reconstructed edge image.

\section{Results}

In this section, we will carry out the numerical simulation of GEGI method for different target objects, and select an aircraft model as the real object for the actual experiment. The original images (numerical simulation), speckle patterns and results image all have resolutions of $128\times 128$ pixels.

\subsection{Numerical simulation results}

For different practical application scenarios, we demonstrate two types of numerical simulation that simultaneously carry out edge detection and imaging.

\begin{figure}[htbp]
\centering
\includegraphics[width=10cm]{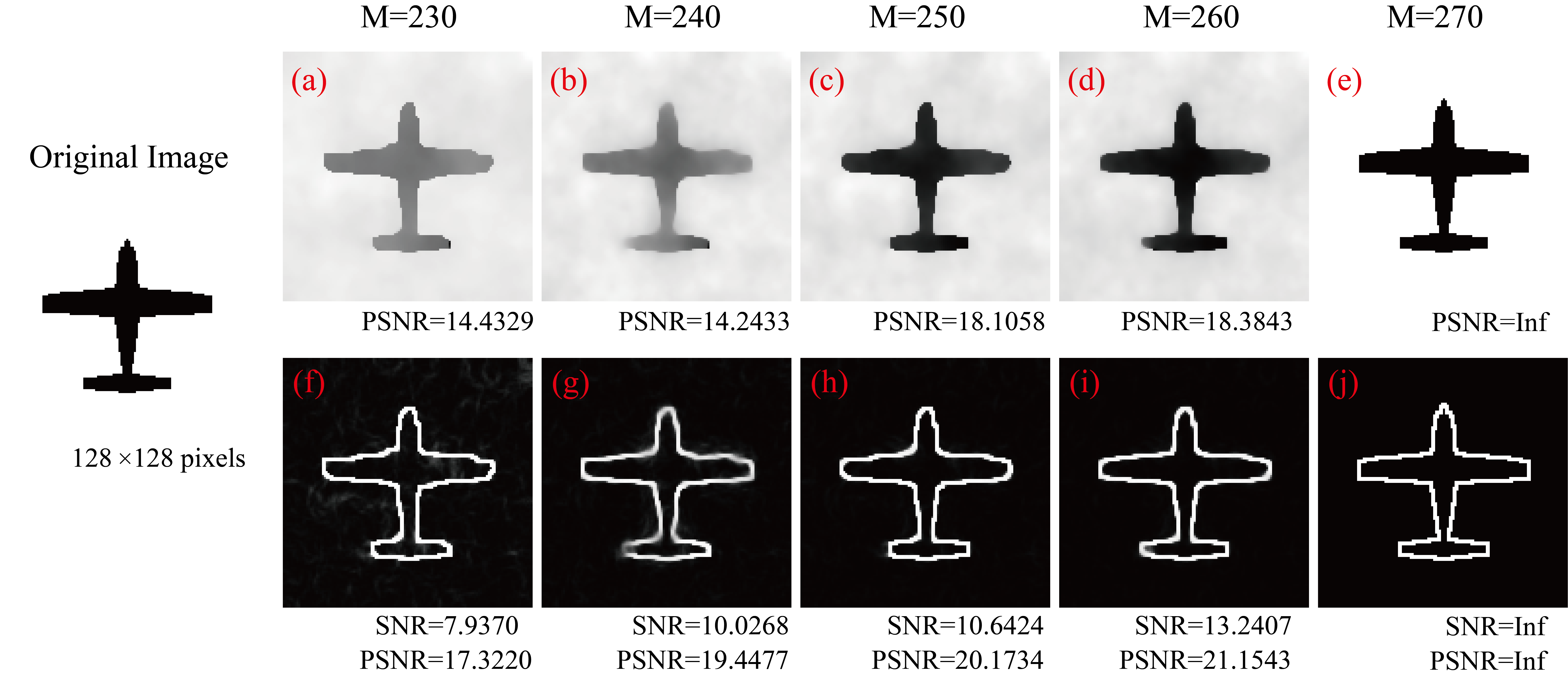}
\caption{(Color online) The numerical simulation results of the aircraft object, where SNRs and PSNRs are presented together.}
\label{air-img}
\end{figure}

\textit{Edge detection in a sparse scene with a large field of view}, e.g., airplanes in the sky, ships in the sea. To simulate this scenario, we use a white background (the pixel value is 1) as the large field of view environment, and the aircraft (the pixel value is 0) as the target object in the scene [see Fig.~\ref{air-img} Original image]. The original image size is $128\times 128$ pixels. With M=230-270 random binary speckle patterns the JIGI results are shown in Fig.~\ref{air-img}, and the SNRs and PSNRs values of the reconstructed images are listed below the corresponding results. With definition in Eq.~(\ref{ngfold}) and PLIR, the simultaneous acquisition of whole image [Fig.~\ref{air-img}(a)-(e)] and global edge [Fig.~\ref{air-img}(f)-(j)] information is realized. With M=230, the blurry image of aircraft is obtained in Fig.~\ref{air-img}(a). However, the corresponding edge image is clearer [as shown in Fig.~\ref{air-img}(f)]. The PSNR of edge image is higher than that of whole image 2.8891 dB. When the number of measurements is increased to 260, the edge and image quality are greatly improved. The SNR and PSNR of the edge are increased to 13.2407 and 21.1543 dB. And the PSNR of the image is increased to 18.3843 dB. Excitedly, with M=270, the results converged to the original image are obtained by joint iteration of PLIR and guided filter [see Fig.~\ref{air-img} (e) and (j)].

\begin{figure}[htbp]
\centering
\includegraphics[width=10cm]{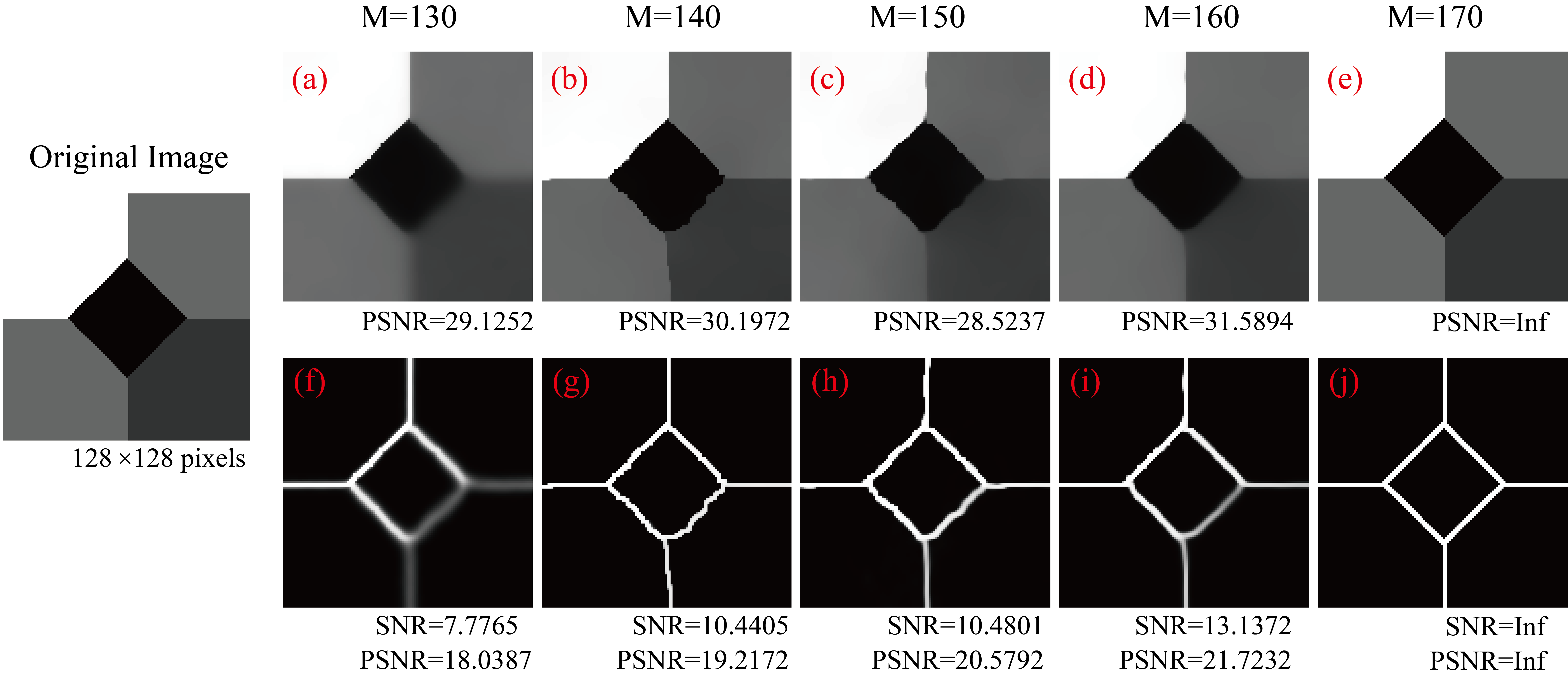}
\caption{(Color online) The numerical simulation results of the simple gray scale object, where SNRs and PSNRs are presented together.}
\label{gray-img}
\end{figure}

\textit{Edge detection in complex scene with gray scale.} The application of edge detection based on JIGI is not only to binary objects, but also to gray scale objects. For the actual application scene, the edge detection of unknown target are more gray scale objects. First of all, a simple gray scale image is treated as the target object which is commonly used for edge detection, as shown in Fig.~\ref{gray-img} Original image. As can be seen from Fig.~\ref{gray-img}, the edge and image information of the simple gray scale objects are effectively acquired, and the SNR and PSNR of the edge are increased to 13.1372 and 21.7232 dB. Similar to the phenomenon in Fig.~\ref{air-img} (e) and (j), when the measurement number is 170, the obtained results converge infinitely to the original image, as shown in Fig.~\ref{gray-img} (e) and (j). The results in Fig.~\ref{gray-img} show that GEGI method can still obtain high quality reconstruction results for simple gray scale object at lower measurement times.

\begin{figure}[htbp]
\centering
\includegraphics[width=10cm]{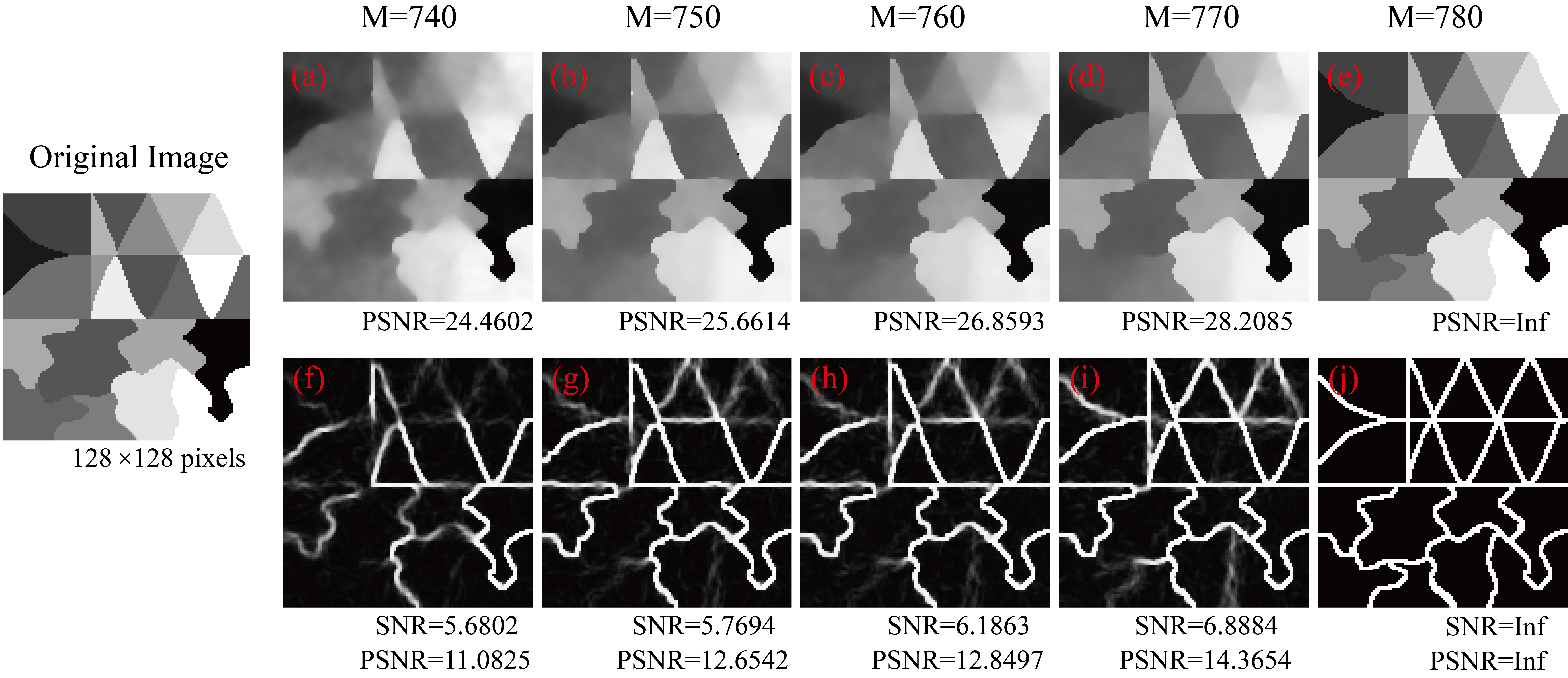}
\caption{(Color online) The numerical simulation results of the complex gray scale object, where SNRs and PSNRs are presented together.}
\label{chaotic-img}
\end{figure}

To further illustrate the effectiveness of JIGI, a gray scale image with more gray levels and higher edge complexity is used and the results are shown in Fig.~\ref{chaotic-img}. The same with the previous simulation results is that when the measurement times of this complex gray scale object is 780, JIGI can still get the reconstruction results which almost identical to the original image, as shown in Fig.~\ref{chaotic-img} (e) and (j). Due to the multiple gray scale and edge complexity, the measurement times of convergence are higher than those in Fig.~\ref{air-img} (M=270) and Fig.~\ref{gray-img} (M=170).

\subsection{Experimental results}

\begin{figure}[htbp]
\centering
\includegraphics[width=12cm]{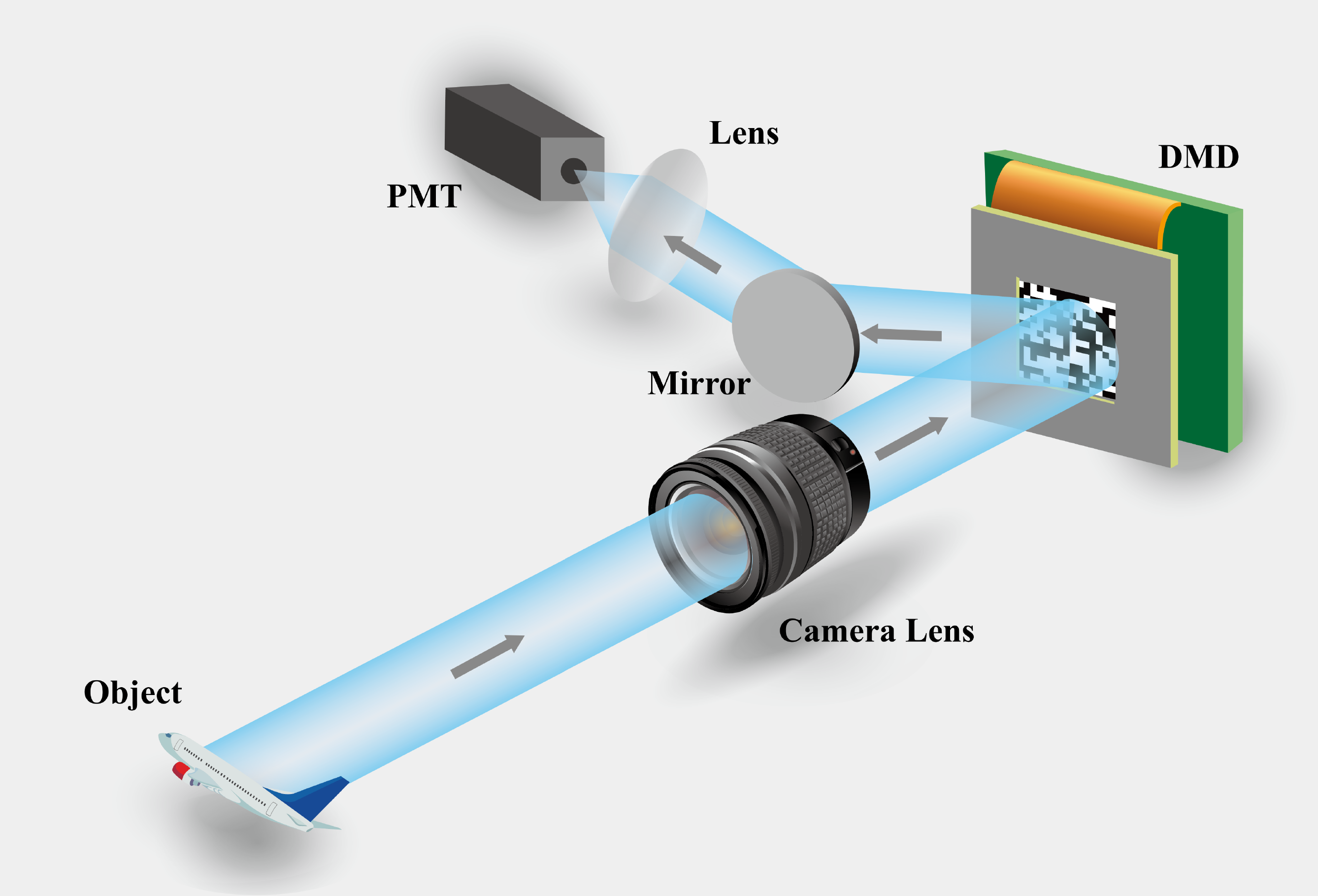}
\caption{(Color online) The experiment system diagram of computational ghost imaging.}
\label{escheme}
\end{figure}

\begin{figure}[htbp]
\centering
\includegraphics[width=10cm]{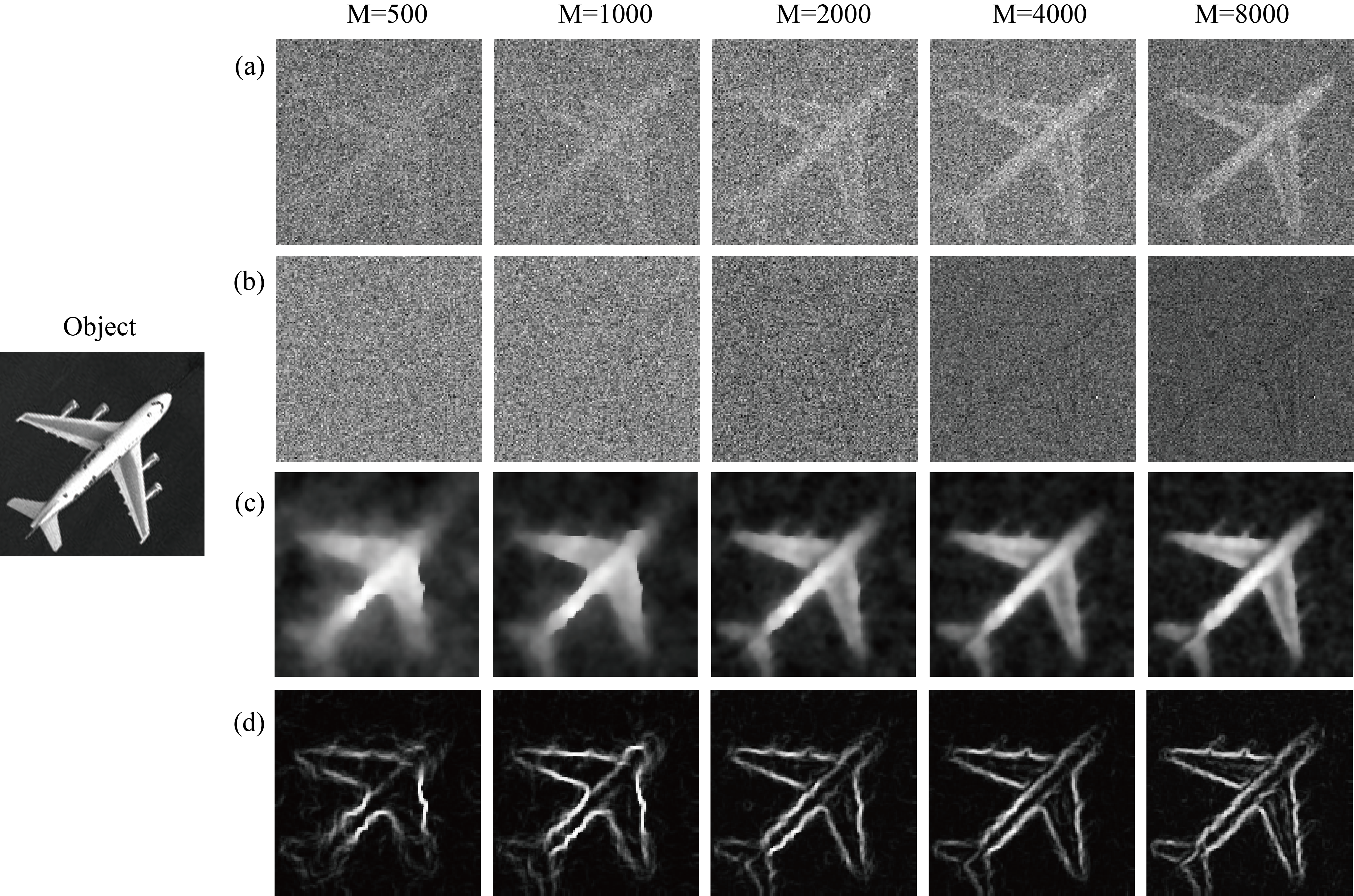}
\caption{Reconstructed images obtained at different measurement times. (a) and (c) are the experimental imaging results of GI and JIGI respectively, (b) and (d) are the edge detection experimental results of GGI and JIGI respectively.}
\label{e-img}
\end{figure}

To demonstrate the feasibility of this JIGI scheme, the actual experiment is conducted. The experiment system configuration is illustrated in Fig.~\ref{escheme}, which includes a camera lens, a DMD, a reflecting mirror, a collecting lens and a photomultiplier tube (PMT). The applied DMD is an excellent device in the scheme for pixel multiplexed modulation and consists of $1024\times 768$ micromirrors, each of which can be switched between two directions of $\pm 12^\circ$, corresponding to 1 and 0. The DMD displayed a preloaded sequence of random speckle patterns ($128~\textrm{pixel}\times 128~ \textrm{pixel}$) at rate of 1K patterns/s. Under ambient illumination (cold white LED), the target object is imaged onto the DMD by the camera lens. A current output type Hamamatsu H10721-01 PMT is placed on one of the reflection orientations to make the measurement of total light signal. The target object is an aircraft model (see Fig.~\ref{e-img} Object) with the size of $20~\textrm{cm}\times 17~\textrm{cm}$ and positioned about 5m away from the DMD.

For comparison, CGI \cite{tcgi}, GGI \cite{ggi} and JIGI experiments are taken for the aircraft model. The imaging results of CGI are shown in Fig.~\ref{e-img}(a). Since the GGI method needs to select the gradient vector of $+45^\circ$ and $-45^\circ$  for edge detection,  the measurement times of PMT detector to obtain global edge image [as shown in Fig.~\ref{e-img}(b)] is triple the number of the CGI random patterns (total $M\times 3$ measurements). According to the experimental results in Fig.~\ref{e-img}(b), GGI's edge detection quality is poor, and the effective edge information can be obtained hardly. However, by using JIGI algorithm, we can acquire a satisfactory image quality, as shown in the illustration of Fig.~\ref{e-img} (c)-(d). Specifically, Fig.~\ref{e-img} (c)-(d) shows five reconstructed images of the aircraft model by using different numbers of patterns. We can see that better quality of reconstructed edge and images with the measurement times increasing. When the measurement times is 500, the imaging and edge detection results of JIGI are obviously better than CGI and GGI, and the edge shape of the aircraft can be roughly distinguished. When the measurement times increases to 1000, the edge information of the aircraft model can be clearly distinguished. As the number of measurements is further increased, the details of the reconstructed image gradually emerge. For example, the engines on both sides of the aircraft have been reconstructed by measuring 8000 times, and the edge and texture feature information is more abundant [as shown in Fig.~\ref{e-img}(d) M=8000]. The results of JIGI experiments verify the feasibility of simultaneously carry out high quality edge detection and imaging.

\section{Conclusion}

In this paper, we have proposed and demonstrated a new edge detection method named joint iteration ghost imaging which uses joint iteration of projected Landweber iteration regularization and guided filtering to realize the high quality edge and whole image information acquisition at the same time. The numerical simulations and experiments show that our JIGI method is validated. Moreover, the proposed method could directly extract the high quality edges in any direction or any GI experimental scheme (e.g. CGI, pseudo-thermal GI, etc.), no matter whether the unknown object is binary or grayscale. We also have compared the performance of CGI and GGI by experiments. The results have showed that the measurement times could be dramatically reduce by using JIGI. As guided filter has the ability of image matting, image super-resolution and haze removal, we believe that edge detection based on JIGI will be valuable in many real applications such as remote sensing, security check and medical imaging \cite{xmi1,xmi2}.

\section{Funding}

This work is supported by the Project of the Science and Technology Department of Jilin Province (Grant No. 20170204023GX); the Special Funds for Provincial Industrial Innovation in Jilin Province (Grant No. 2018C040-4, 2019C025); the Science Foundation of Education Department of Jilin Province (Grant No. 2019LY508L35).




\end{document}